# Ferroelastic Altermagnetism

Rui Peng[1], Shibo Fang[1], Pin Ho[2], Tong Zhou[3], Junwei Liu[4, *], Yee Sin Ang[1, †]

[1] Science, Mathematics and Technology (SMT) Cluster, Singapore University of Technology and Design (SUTD), Singapore 487372, Singapore.
[2] Institute of Materials Research and Engineering (IMRE), Agency for Science, Technology and Research (A*STAR), Singapore 138634, Singapore.
[3] Eastern Institute for Advanced Study, Eastern Institute of Technology, Ningbo, Zhejiang 315200, China
[4] Hong Kong University of Science and Technology, Hong Kong, China SAR.

Emails: *liuj@ust.hk, †yeesin_ang@sutd.edu.sg

## Abstract

Combination of altermagnetism and ferroic orders, such as ferroelectric switchable altermagnetism [*Phys. Rev. Lett.* **134**, 106801 (2025) and *Phys. Rev. Lett.* **134**, 106802 (2025)], offers a powerful route to achieve nonvolatile switching of altermagnetic spin splitting. In this work, by synergizing altermagnetism and ferroelasticity, we propose the concept of *ferroelastic altermagnets* in which the ferroelastic crystal reorientation can drive multistate nonvolatile switching of the altermagnetic spin splitting via *altermagnetoelastic effect*. Using monolayers $RuF_4$ and $CuF_2$ as material candidates, we demonstrate 2-state and 3-state altermagnetic spin splitting switching as driven by ferroelastic strain states. Transport calculation shows that multistate spin conductivities can be ferroelastically encoded in an ferroelastic altermagnet, thus suggesting the potential of ferroelastic altermagnetic as nonvolatile nanomechanical spin switches. The proposed concept of ferroelastic altermagnetism enriches the emerging landscape of multiferroic altermagnetism and shall pave a way towards straintronic-spintronic device applications.

# 1. Introduction

Together with ferromagnet (FM) and conventional antiferromagnet (AFM), altermagnet represents an emerging new family of collinear magnetic materials [1-5]. Altermagnets exhibit zero net magnetization like conventional AFM, but host nonrelativistic band spin splitting like FM. Such FM-AFM dichotomy endows altermagnets with the advantages of AFM such as robustness against external fields and compatible with ultrahigh-speed device operation while still exhibiting broken spin degeneracy of FM, which are critical for spin manipulation and information processing [6-21]. Altermagnets exhibit a plethora of intriguing physical phenomena such as spin current generation [1-2], spin Hall effect [6], spin Nernst effect [7], anomalous Hall effect [5], and tunneling magnetoresistance effects [6], which can be harnessed for various device applications. A large variety of altermagnets have been theoretically proposed [3-4,17,22], and some of them have been experimentally confirmed [23-26].

The physics of altermagnets can be further enriched by *multiferroicity*. Multiferroics are singular materials that exhibit two or more ferroic orders among (anti)ferromagnetism, (anti)ferroelectricity and (anti)ferroelasticity [27-35]. For instance, the coupling between ferroelectricity and magnetism yields *magnetoelectric effect* where ferroelectric polarization and magnetization are coupled and can be mutually switched. Recent studies have unveiled the altermagnetic counterpart of magnetoelectric effect, namely the *altermagnetoelectric effect* [**Fig. 1(a)**] in which the altermagnetic spin splitting can be tuned by various forms of ferroelectric switching [36-43], thus unveiling a route towards electrical-based nonvolatile switching of altermagnetism. The multiferroic integration of ferroelectricity and altermagnetism immediately raises the following question: Can altermagnetism be integrated with *ferroelasticity* [44-48] – *magnetoelastic effect* – to achieve nonvolatile *nanomechanical* deformation-induced switching of the altermagnetic spin splitting?

Here we propose the concept of *ferroelastic altermagnetism* in which altermagnetism and ferroelasticity are synergized. As ferroelastic switching is equivalent to a rotation operation on the lattice, such lattice rotation changes the spin-momentum coupling in a ferroelastic altermagnetism, thus enabling an *altermagnetoelastic effect* in which the altermagnetic spin splitting can be mechanically switched [**Figs. 1(b,c)**]. Ferroelastic altermagnetism represents a *mechanical* multiferroic counterpart of the recently proposed ferroelectric altermagnetism [36-40]. Using monolayers RuF$_4$ [49] and CuF$_2$ as proof-of-concept, 2-state and 3-state nonvolatile switching of the altermagnetic spin splitting are demonstrated, respectively, via first-principles calculations. We further show that the multistate ferroelastic switching lead to nonvolatile switchable spin transport. These findings reveal a previously unexplored mechanism to achieve nonvolatile strain-based altermagnetic switching that enables *multistate* nonvolatile information encoding. This concept of ferroelastic altermagnet shall serve as a harbinger of straintronic-spintronic devices uniquely enabled by the union of altermagnetism and ferroelasticity.

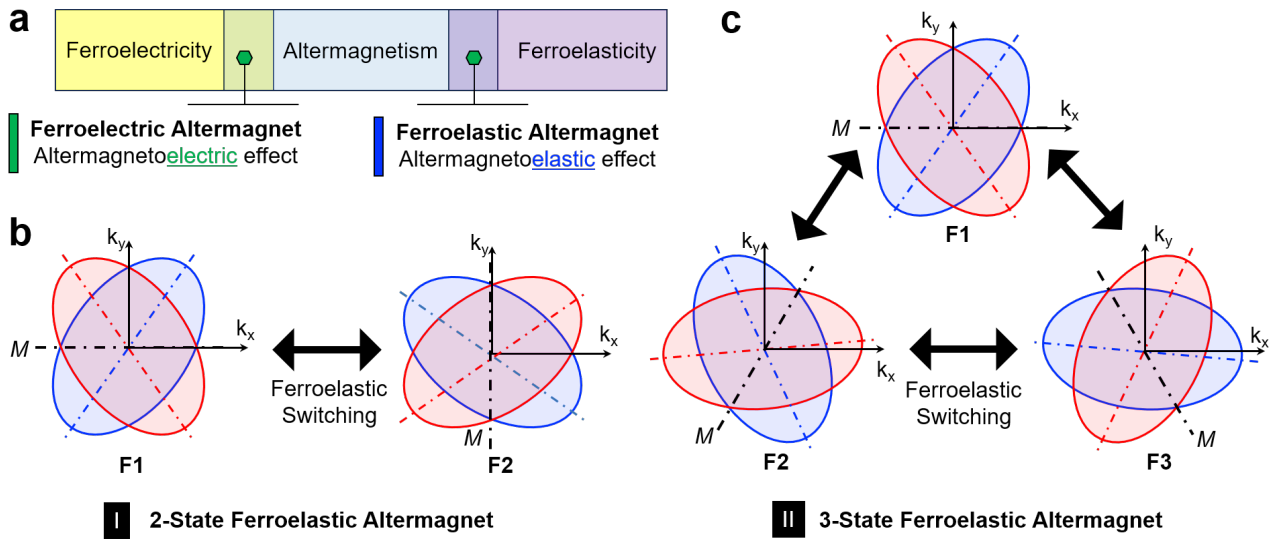

**Fig.1. Concept of ferroelastic altermagnetism.** (a) Ferroelastic altermagnet arises in a material where altermagnetism and ferroelasticity are simultaneously present. Schematic diagram of (b) 2-state and (c) 3-state nonvolatile mechanical switching of the altermagnetic spin splitting two-dimensional ferroelastic altermagnets. M is the mirror symmetry axis of the systems.

## 2. Results and Discussion

*2.1 Ferroelastic-driven altermagnetic spin switching and material candidates*

Two-dimensional (2D) materials can be classified according to layer groups [50], and those belonging to layer groups $L_8$–$L_{48}$ may host ferroelasticity due to their rectangular 2D Bravais lattices. Altermagnetic monolayers within these layer groups may thus exhibit coexisting ferroelasticity and altermagnetism [18]. To illustrate the concept of ferroelastic altermagnetism, we use monolayers RuF$_4$ and CuF$_2$ (layer group $L_{17}$) as a proof-of-concept material platform for achieving 2-state and 3-state ferroelastic-driven altermagnetic spin switching. Monolayers RuF$_4$ and CuF$_2$ can potentially be mechanically exfoliated from bulks RuF$_4$ and CuF$_2$ due to their layered structures [51-52]. As shown in **Fig. 2(a)**, the crystal lattice of monolayer RuF$_4$ has a rectangular lattice with a space group of *P2$_1$/c*. The Ru atoms are encapsulated in a distorted RuF$_6$ octahedron, and each Ru atom is connected to six F atoms. The optimized lattice constants of monolayer RuF$_4$ is *a* = 5.42 Å and *b* = 5.08 Å. The crystal structure of monolayer CuF$_2$ shares the same space group *P2$_1$/c* with monolayer RuF$_4$ [**Fig. 2(b)**]. Also, the Cu atoms are encapsulated in a distorted CuF$_6$ octahedron, and each Cu atom is connected to six F atoms. The difference is that the adjacent RuF$_6$ octahedrons are connected via the corner-sharing manner. The optimized lattice constants of monolayer CuF$_2$ is *a* = 5.32 Å and *b* = 3.56 Å. The stability of monolayers RuF$_4$ and CuF$_2$ are confirmed by phonon and *ab initio* molecular dynamics simulations (Fig. S1 [53]).

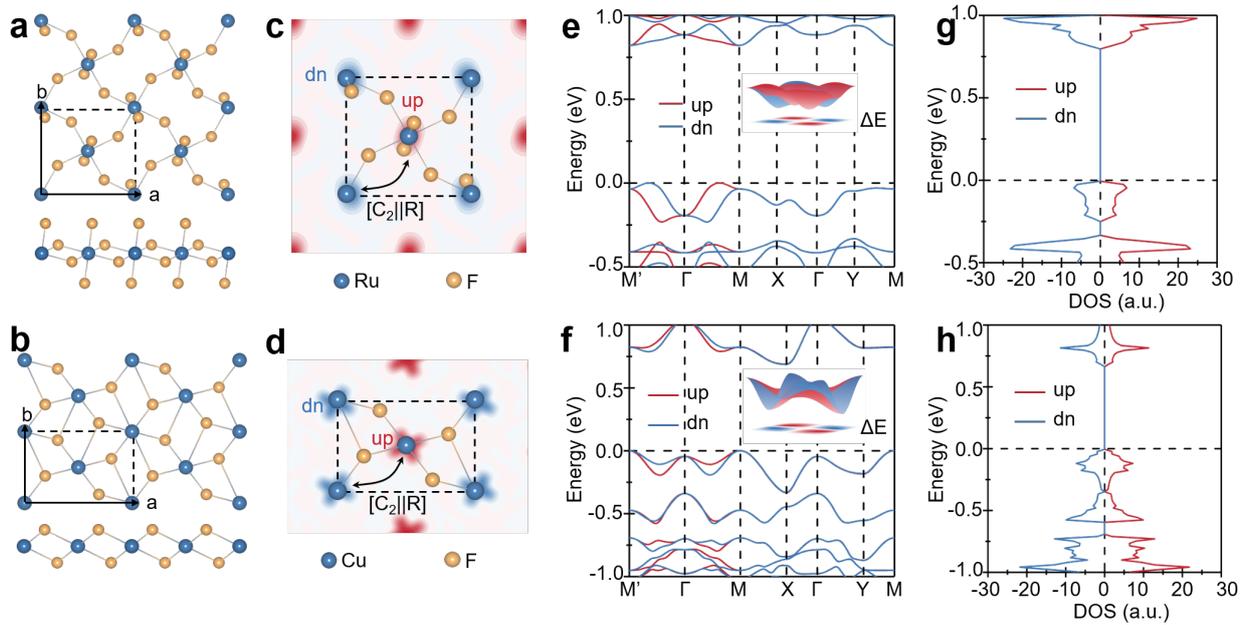

**Fig. 2. Crystal and electronic structures of 2D ferroelastic altermagnets.** Crystal structures of (a) RuF$_4$ and (b) CuF$_2$ from top and side views. Spin charge densities of monolayers (c) RuF$_4$ and (d) CuF$_2$. Red and blue shading represent spin up and down, respectively. Band structures of monolayers (e) RuF$_4$ and (f) CuF$_2$. Insets are two highest valence bands, and alternating spin splitting ΔE between them in the hole Brillouin zone. Density of states of monolayers (g) RuF$_4$ and (h) CuF$_2$. Fermi level is set to zero in (e) to (h).

Under the octahedral crystal field, the d orbitals split into t$_{2g}$ triplets and e$_g$ doublets. In monolayer RuF$_4$, Ru atoms lose four electrons and transform into Ru$^{4+}$ ions with four 4d electrons. Two of them occupy one t$_{2g}$ orbital, one of them occupy another t$_{2g}$ orbital, and one of them occupy the remaining $t_{2g}$ orbital, giving rise to a local magnetic moment of 2 $\mu_B$. The inequivalent distribution of the electrons on the three $t_{2g}$ orbitals leads to the compressed Jahn–Teller distortions of the crystal structure, which further splits the $t_{2g}$ and $e_g$ orbitals (Fig. S2 [53]). In monolayer $CuF_2$, Cu atoms lose two electrons and transform into $Cu^{2+}$ ions with nine 3d electrons. Six of them occupy the three $t_{2g}$ orbitals, two of them occupy one $e_g$ orbital, and one of them occupy the other $e_g$ orbital, giving rise to a local magnetic moment of 1 $\mu_B$. The inequivalent distribution of the electrons on the two $e_g$ orbitals leads to the elongated Jahn–Teller distortions of the crystal structure, making $t_{2g}$ and $e_g$ orbitals split in a different way with those in monolayer $RuF_4$ (Fig. S2 [53]). The calculated magnetic moments on Ru and Cu atoms are 1.67 and 0.64 $\mu_B$, respectively, which is consistent with the above analysis.

The ground states of monolayers $RuF_4$ and $CuF_2$ are determined by comparing the energy among three magnetic configurations: FM, Néel antiferromagnetic (AFM-Néel), and stripe antiferromagnetic (AFM-stripe) (Fig. S3 [53]). It is found that the energies of FM and AFM-stripe are 30.28 (7.47) and 11.69 (4.27) meV/unit cell higher than the AFM-Néel in monolayer $RuF_4$ ($CuF_2$), respectively, thus indicating AFM-Néel is the energetically most favorable state for both $RuF_4$ and $CuF_2$. The spin charge densities of AFM-Néel $RuF_4$ and $CuF_2$ are shown in **Figs. 2(c-d)**. After considering magnetism, the two opposite-spin sublattices in $RuF_4$ and $CuF_2$ can only be transformed to each other by screw or glide symmetry, which is typical for altermagnets.

Monolayer $RuF_4$ has an indirect band gap of 0.82 eV, with CBM located at M/M' point and VBM located along Γ-M/M' line [**Fig. 2(e)**]. The spin-up and spin-down states are degenerate along M-X-Γ-Y-M, while an alternating spin splitting appears along M'-Γ-M, forming a spin-momentum locking physics. We also plotted the two highest valence bands and the amplitude of the spin splitting ΔE between them over the entire Brillouin zone. The spin-momentum locking physics can also be observed at generic k points. The amplitude of the spin splitting reaches up to 205 meV. For monolayer $CuF_2$, its indirect band gap is 0.69 eV, with CBM located at X point and VBM located at M/M' point [**Fig. 2(f)**]. Similarly, the spin-up and spin-down states are degenerate along M-X-Γ-Y-M, while an alternating spin splitting appears along M'-Γ-M. The amplitude of the spin splitting reaches up to 98 meV. The spin-up and spin-down density of states of $RuF_4$ and $CuF_2$ are equal [**Figs. 2(g-h)**].

## 2.2 Symmetry analysis

The underlying mechanism of the spin-momentum locking in monolayers RuF$_4$ and CuF$_2$ can be elucidated through symmetry analysis. When omitting the nonmagnetic F atoms, the systems have PT and t$_{1/2}$T symmetry, thus guaranteeing that the band structures are spin-degenerate over the entire Brillouin zone. The F atoms in the noncentrosymmetric sites break the PT and t$_{1/2}$T symmetry, yielding spin splitting in the band structures. The space group of monolayers RuF$_4$ and CuF$_2$, *P2$_1$/c*, contains four symmetry operations: {E, P, t$_{1/2}$M$_y$, t$_{1/2}$C$_{2y}$}. The former two connects the same-spin sublattices, while the latter two connects the opposite-spin sublattices. Considering the crystal and spin rotation symmetries remain decoupled when considering AFM coupling in the absence of spin-orbital coupling, the symmetry connecting two opposite-spin sublattices can be represented as [C$_2$||R], where the transformations on the left (right) of the double vertical bar act only on the spin (real) space. When the [C$_2$||t$_{1/2}$M$_y$] and [C$_2$||t$_{1/2}$C$_{2y}$] is operated on monolayer CuF$_2$, respectively, we obtain:

$$[C_2||t_{1/2}M_y]\mathrm{E}(k_x, k_y, \uparrow) = \mathrm{E}(k_x, -k_y, \downarrow)$$

$$[C_2||t_{1/2}C_{2y}]\mathrm{E}(k_x, k_y, \uparrow) = \mathrm{E}(-k_x, k_y, \downarrow)$$

(1)

Consequently, we have

$$\mathrm{E}(k_x, k_y, \uparrow) = \mathrm{E}(k_x, -k_y, \downarrow)$$

$$\mathrm{E}(k_x, k_y, \uparrow) = \mathrm{E}(-k_x, k_y, \downarrow)$$

(2)

This explains the emergence of alternating spin-split bands along M'-Γ-M and at generic k points. We also have

$$\mathrm{E}(k_x, 0, \uparrow) = \mathrm{E}(k_x, 0, \downarrow)$$

$$\mathrm{E}(0, k_y, \uparrow) = \mathrm{E}(0, k_y, \downarrow)$$

(3)

After considering the Bloch theorem, the bands are spin-degenerate when $k_x = m\pi$ and $k_y = n\pi$, thus explaining the spin-degenerate bands along high-symmetry paths M-X-Γ-Y-M.

## 2.3 Multistate nonvolatile ferroelastic-altermagnetic spin splitting switching

Interestingly, $RuF_4$ and $CuF_2$ are also ferroelastic monolayers. $RuF_4$ has two energy-equivalent states F1 (a > b) and F2 (a < b), and one paraelastic state P (a = b) [**Fig. 3(a)**]. The ferroelastic switching from F1 to F2 is equivalent to a π/2 rotation of the lattice. In contrast, $CuF_2$ has three ferroelastic states. For convenience, we choose the deformed hexagon to be the structure unit. The three energy-equivalent ferroelastic phases can be described as F1 ($a_1 > a_2/a_3$), F2 ($a_2 > a_1/a_3$) and F3 ($a_3 > a_1/a_2$), where $a_1$, $a_2$ and $a_3$ are the diagonal lines of the deformed hexagon [**Fig. 3(b)**]. The paraelastic phase P is the T-phase $CuF_2$ with $a_1 = a_2 = a_3$. The ferroelastic switching from F1 to F2 (F3) is equivalent to a 2π/3 (4π/3) lattice rotation. The ferroelastic reversible strain of $RuF_4$ and $CuF_2$, which are defined as (a/b - 1) × 100% and ($a_i/a_j$ - 1) × 100%, are calculated to be 6.8% and 11.4%, respectively, which are larger than those of SnS (4.9%) and SnSe (2.1%) [44].

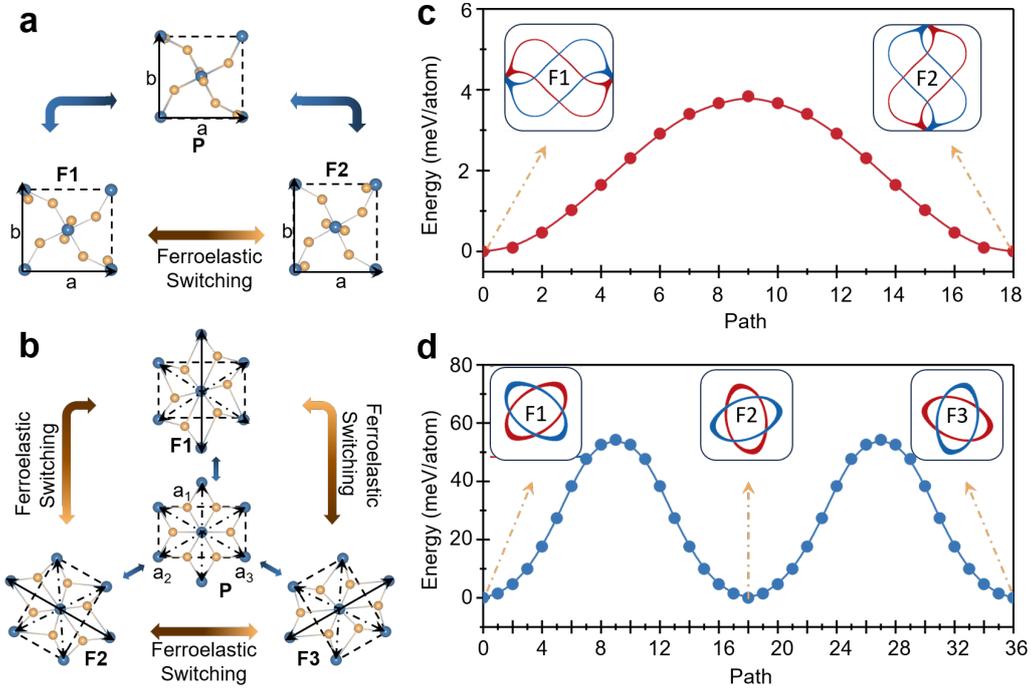

**Fig. 3. Ferroelastic switching in $RuF_4$ and $CuF_2$.** Schematic diagram of 2-state and 3-state ferroelastic switching in (a) monolayer $RuF_4$ and (b) monolayer $CuF_2$, respectively. Here $a_1$, $a_2$ and $a_3$ are three diagonal lines of the deformed hexagon of $CuF_2$. Energy profiles of the ferroelastic transition for (c) $RuF_4$ and (d) $CuF_2$ as a function of step number in nudged elastic band (NEB) calculations. Insets show the spin-momentum locking in different ferroelastic phases at −0.1 eV below the Fermi level.

We simulate the ferroelastic switching process of RuF$_4$ and CuF$_2$ using the nudged elastic band (NEB) method. The overall switching barrier of RuF$_4$ and CuF$_2$ are 4 and 54 meV/atom [**Figs. 3(c-d)**], respectively, which is lower than those of phosphorene (200 meV/atom) [44] and borophane (100 meV/atom) [45]. The low switching barrier suggests that monolayers RuF$_4$ and CuF$_2$ are potentially compatible with fast and low-energy ferroelastic switching operation. Upon ferroelastic switching, the reoriented lattice changes spin-momentum locking fashion, thus realizing multistate nonvolatile switching of the altermagnetic spin splitting [see insets in **Figs. 3(c-d)**].

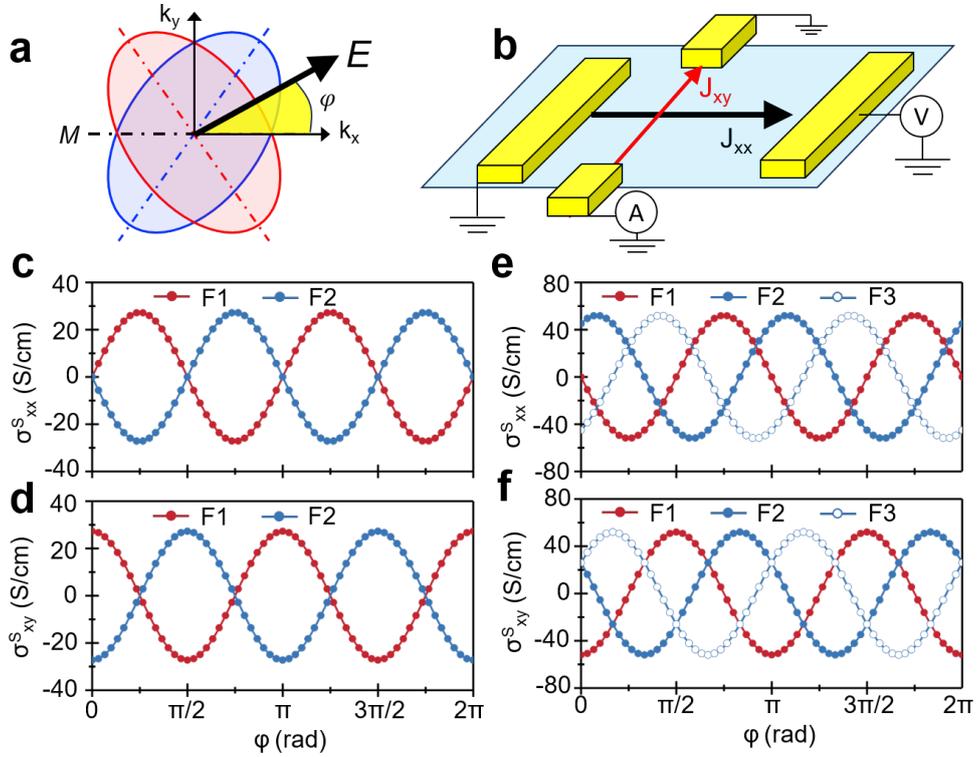

**Fig.4. Transport properties in RuF$_4$ and CuF$_2$.** (a) Schematic diagram of the direction of the external in-plane electric field. (b) Schematic diagram of a four-terminal Hall-bar-like device. (c) Longitudinal and (d) transverse spin conductivities of monolayer RuF$_4$ as a function of φ. (e) Longitudinal and (f) transverse spin conductivities of monolayer CuF$_2$ as a function of φ. Fermi level is set to -0.1 eV in (c-f).

In altermagnets, the spin-polarized current can be generated by an external in-plane electric field due to the anisotropic spin splitting [1,2,40]. In ferroelastic altermagnets, the spin transport current can be mechanically switched via different ferroelastic states. We calculate the transport current in RuF$_4$ and CuF$_2$ when driven by an external field [53]. The longitudinal and transverse spin conductivities exhibit an angular dependence on φ with a period of π, where φ is the direction of the in-plane electric field relative to *x* axis [**Fig. 4(a)**] and φ can be experimentally controlled by changing the relative angle between the source-drain electrode axis [see **Fig. 4(b)** for a device schematic] and the crystal structure of ferroelastic altermagnet sample. Upon a ferroelastic switching, the lattice re-orientation changes the relative orientation between the altermagnetic spin-split band structure and the electric field, thus leading to spin current switching effect. In $RuF_4$, the carrier transport under F1 and F2 states exhibits equal but opposite transverse and longitudinal spin conductivities [**Figs. 4(c-d)**]. The 2-state ferroelastic switching between F1 and F2 states thus lead to spin reversal. For $CuF_2$, the spin conductivities exhibit even richer switching effect via the three ferroelastic states F1, F2 and F3. For instance, when the electric field is directed at φ=0, the longitudinal spin conductivities under F1, F2 and F3 states can be switched between positive, zero and negative magnitudes, respectively [**Figs. 4(e-f)**]. The ferroelastic altermagnetic broaden the application scenario of altermagnets, enabling nonvolatile spin information encoding and hybrid straintronic-spintronic device application.

## 3. Conclusion

In summary, we have proposed the concept of ferroelastic altermagnets that synergizes ferroelasticity with altermagnetism. The switchable ferroelastic states in ferroelastic altermagnets drive multistate nonvolatile switching of the altermagnetic spin splitting, leading to an *altermagnetoelastic effect*. Using monolayers $RuF_4$ and $CuF_2$ as proof-of-concept, 2-state and 3-state nonvolatile switching of the altermagnetic spin splitting are demonstrated. We further show that the multistate ferroelastic phase transitions are accompanied by a switching of the spin current associated with the altermagnetic spin splitting. We note that the ferroelastic altermagnetism can also be achieved in other 2D ferroelastic altermagnets such as the two-state ferroelastic monolayers of $VF_4$ [11,13,54,55] and $OsF_4$ [13], and the three-state ferroelastic monolayer $AgF_2$ [13,56,57].

## 4. Experimental Section

First-principles calculations are performed based on density functional theory (DFT) [58] as implemented in Vienna ab initio simulation package (VASP) [59]. Exchange-correlation interaction is described by the Perdew-Burke-Ernzerhof (PBE) parametrization of generalized gradient approximation (GGA) [60]. Structures are relaxed until the force on each atom is less than 0.01 eV/Å. The cutoff energy and electronic iteration convergence criterion are set to 600 eV and $10^{-6}$ eV, respectively. To sample the 2D Brillouin zone, Monkhorst–Pack (MP) k-grid messes [61] of 7 × 7 × 1 and 5 × 9 × 1 are used for monolayers $RuF_4$ and $CuF_2$, respectively. To avoid the interaction between adjacent layers, a vacuum space of 20 Å is added. Phonon calculations are carried out using the PHONOPY code with a 4 × 5 × 1 and 4 × 4 × 1 supercell for monolayers $RuF_4$ and $CuF_2$, respectively [62]. Ab initio molecular dynamics (AIMD) simulations are performed at 300 K for 5 ps with a time step of 1 fs with a 4 × 5 × 1 and 4 × 4 × 1 supercell for monolayers $RuF_4$ and $CuF_2$, respectively. Ferroelastic switching barrier is obtained by nudged elastic band (NEB) method [63]. The spin-resolved transport properties are calculated using a housing-made code, in which the electron energy and electron group velocity are evaluated from the Wannier-based tight binding Hamiltonian [64].

## Acknowledgement

This work is supported by the Singapore Ministry of Education Academic Research Fund (AcRF) Tier 2 under the Project ID MOE-000834-00 (Proposal ID T2EP50224-0006). P.H. acknowledges the supports by the RIE2025 Manufacturing, Trade and Connectivity (MTC) Individual Research Grants (Grant No. M23M6c0101) and (Grant No. M24N7c0086), administered by A*STAR. T.Z. acknowledges the supports by the National Natural Science Foundation of China (12474155) and the Zhejiang Provincial Natural Science Foundation of China (LR25A040001). The computational work for this article was performed on resources of the National Supercomputing Centre, Singapore (https://www.nscc.sg).

## Conflict of Interest

The authors declare no conflict of interest.

## Data Availability Statement

The data that support the findings of this study are available in the supplementary material of this article.

## Keywords

Altermagnetism, Ferroelasticity, Multiferroicity, First-principles calculation, Nonvolatile switching